\documentclass[preprint2]{aastex}          
\newcommand{\cf}{{\em cf. }}
 
\newcommand{\eg}{{\em e.g. }}

\begin{document}
%
\title{Mass-luminosity relation for FGK main sequence stars: \\metallicity and age contributions}

\shorttitle{Mass-luminosity relation for FGK main sequence stars}
\shortauthors{RIcardo Gafeira}

\author{Ricardo Gafeira \altaffilmark{1}}
\and \author{Carlos Patacas\altaffilmark{2}}
\and
\author{Jo\~ao Fernandes\altaffilmark{1,3}}

\altaffiltext{1}{Centro de Geof\'isica da Universidade de Coimbra, Observat\'orio Astron\'omico da Universidade de Coimbra, Santa Clara, Coimbra, Portugal.}
\altaffiltext{2}{Laborat\'orio de Instrumenta\c c\~ao e F\'isica Experimental de Part\'iculas, Departamento de F\'isica, Universidade de Coimbra, Coimbra, Portugal}
\altaffiltext{3}{Departamento de Matem\'atica e Observat\'orio Astron\'omico da Universidade
de Coimbra, Santa Clara, Coimbra, Portugal.}

\begin{abstract}
The stellar mass-luminosity relation (MLR) is one of the most famous empirical ``laws", discovered in the beginning of the 20th century. MLR is still used to estimate stellar masses for nearby stars, particularly for those that are not binary systems, hence the mass cannot be derived directly from the observations. It's well known that the MLR has a statistical dispersion which cannot be explained exclusively due to the observational errors in luminosity (or mass). It is an intrinsic dispersion caused by the differences in age and chemi\-cal composition from star to star. In this work we discuss the impact of age and metallicity on the MLR. Using the recent data on mass, luminosity, metallicity, and age for 26 FGK stars (all members of binary systems, with observational mass-errors $\leq3\%$), including the Sun, we derive the MLR taking into account, sepa\-rately, mass-luminosity, mass-luminosity-metallicity, and mass-luminosity-metallicity-age. Our results show that the inclusion of age and metallicity in the MLR, for FGK stars, improves the individual mass estimation by 5\% to 15\%.\\
\end{abstract}


\section{Introduction}\label{int}

Mass is a fundamental stellar parameter with a crucial impact on 
the evolution and the internal structure of stars, and in other astrophysical situations (\eg
the initial mass function, the mass-luminosity relation, the study of
extra-solar planets, the definition of the star-brown dwarf limit, etc) necessary to study stellar populations.  
Stellar masses can be accurately determined only for visual binaries with known
orbital elements and distances, for detached double-lined eclipsing
binaries, and for resolved spectroscopic binaries. The latter can provide even accuracies of $1\%$  to 3$\%$ \citep{2010A&ARv..18...67T}. 
Unfortunately, the number of
stars for which accurate stellar masses are available is less than 200
\citep{2010A&ARv..18...67T, 2004ApJ...604..741H}. 
Therefore, the estimation of mass for single stars makes use of indirect methods: either observational 
--- as the use of surface gravity obtained from the detailed spectroscopic
analysis \citep{2004A&A...415.1153S} --- or by means of comparison between observations and  theoretical
stellar models \citep{2009A&A...501..941H} or asteroseismology \citep{2012arXiv1201.5966M}. 
However,
on those estimations the mass-luminosity relation (MLR) is particularly useful
as, in principle, it is purely observational (hence, model-independent) and
requires only the knowledge of luminosity (easily determined for nearby
stars). 
This fundamental relation was definitively established by \cite{1924MNRAS..84..308E} 
triggered by previous works \citep[for the
historical MLR overview, please see][]{Lecchini}.
During the last 20 years, thanks to the observational improvements (mainly in parallax and infrared
photometry) several works have been published on
the MLR subject. These works have in common the derivation of formulae accounting for 
the mass-luminosity dependence for a large range of masses: from the brown
dwarf mass limit until O and A spectral type stars. 
We can find polynomial
fits between the mass and luminosity logarithm (or the visual
magnitude) to be linear \citep{1983Ap&SS..96..125C}, parabolic \citep{1993AJ....106..773H}, 
cubic or higher \citep{2002AJ....124.2721R}. 
For an overview, see \cite{2010ChA&A..34..277X}. 
Other than the derived monotonic relations, all the authors
point out that the observational MLR shows a dispersion that
cannot be explained by the observational errors neither in mass nor
luminosity. This is well summarized by \cite{2010A&ARv..18...67T}: 
{\it The significant
effects of both stellar evolution and abundance differences are well seen
in the close-up of the deceptively tight mass-luminosity relation of Fig.
5 that we show in Fig. 6. Making the error bars visible highlights the
fact that the scatter is highly significant and not due to observational
uncertainties}. 
In spite of all the works carried out during the last years on the analysis of age and chemical composition contribution to the MLR \citep[\eg][]{2000MNRAS.315..543H}, 
a MLR involving simultaneously mass, age, chemical composition, and luminosity was never established.
As far as we know, only \cite{2005A&A...442..635B} quoted empirical formulae where the visual
magnitude is dependent both on mass and on metallicity ([Fe/H]) but just for M stars. 
The age was not considered. 
We point out also a kind of mass-luminosity relation derived by \cite{2010A&ARv..18...67T}
where they performed a fit to mass expressed by a polynomial in effective
temperature, gravity, and metallicity. 
However, in this case, age and luminosity are not
explicitly used. 
The main goal of our work is to discuss the
metallicity and age contribution to the MLR. 
For that, we use the data for FGK stars from \cite{2010A&ARv..18...67T} for which
individual values of stellar mass, luminosity, metallicity, and age are available: this gives 13 binary
systems. 
The mass estimations of these binary components have an
accuracy better than $3\%$. This work is organized as follows: i) in Section \ref{mas}, we derive the empirical relation
for the mass as function of luminosity, metallicity, and age; ii) in Section \ref{dis} we
discuss the results and draw the conclusions.

\section{Mass-luminosity-metallicity-age relation: fits and results}\label{mas}
Both theory and observations indicate that the Sun was 20\% to 30\% fainter in ZAMS than today \citep{2003ApJ...583.1024S}.
However, the solar mass remains basically the same during the solar evolution.
Therefore: same mass, different luminosity. 
The impact of chemical composition in the MLR is less clear. But, it can be shown, using simple stellar homological relations, 
that metallicity can have an impact of 0.25$M_\odot$ for solar-like stars (see Section \ref{annex}). 
Consequently, we aim to discuss the impact of metallicity and age on the MLR. 
It is well known that age has a strong impact on the luminosity of a star. In order to accomplish our work, we have selected 13 binary systems
with known mass, luminosity, metallicity, and age for each binary member, taken from the recent compilation of \cite{2010A&ARv..18...67T},
where components are all FGK main sequence stars: the mass ranges from 0.8 to 1.45$M_\odot$; the metallicity from -0.6 to 0.4 dex; and the age from 0.6 to 7.0 Gyr. 
The choice to restrict our analysis to FGK main sequences stars lies on the age determinations. 
Stellar age is currently determined by means of theoretical models and/or isochrones. 
These stellar models are based on the input physics to describe the evolution of the Sun. 
This approach seems to be suitable to reproduce FGK stars, because their global parameters are well recovered 
by means of theoretical models \citep[see][]{lebreton1999, lebreton2008}. 
So, we expect that the ages of our sample are reliable. As the errors on age are not given by \citet{2010A&ARv..18...67T} 
we assume a typical error of $20\%$. Table \ref{data} shows the observations of the selected stars. The Sun is also shown. 

 \begin{table*}[H]
 \small
 \caption{Observational data from \citet{2010A&ARv..18...67T}.} 
  \label{data}
 \begin{tabular}{ccccccccc}
 \tableline  
Name & Log $L/L_\odot$ & error &[Fe/H]& error & $Stellar Age /Age_\odot$&  error  & Mass & error  \\ \hline
V570\_Per A& 0.6580 & 0.0230 & 0.0200 & 0.0300 & 0.1348 & 0.0270 & 1.4466 & 0.0086 \\ \hline
V570\_Per B& 0.5050 & 0.0180 & 0.0200 & 0.0300 & 0.1348 & 0.0270 & 1.3471 & 0.0081 \\ \hline
CD\_Tau A& 0.6320 & 0.0160 & 0.0800 & 0.1500 & 0.6757 & 0.1351 & 1.4420 & 0.0160 \\ \hline
CD\_Tau B& 0.5220 & 0.0170 & 0.0800 & 0.1500 & 0.6757 & 0.1351 & 1.3680 & 0.0160 \\ \hline
AD\_Boo A& 0.6400 & 0.0330 & 0.1000 & 0.1500 & 0.3387 & 0.0677 & 1.4136 & 0.0088 \\ \hline
AD\_Boo B& 0.2780 & 0.0350 & 0.1000 & 0.1500 & 0.3387 & 0.0677 & 1.2088 & 0.0056 \\ \hline
VZ\_Hya A& 0.4800 & 0.0390 & -0.2000 & 0.1200 & 0.3387 & 0.0677 & 1.2713 & 0.0087 \\ \hline
VZ\_Hya B& 0.2410 & 0.0420 & -0.2000 & 0.1200 & 0.3387 & 0.0677 & 1.1459 & 0.0059 \\ \hline
V505\_Per A& 0.4270 & 0.0210 & -0.1200 & 0.0300 & 0.3387 & 0.0677 & 1.2719 & 0.0072 \\ \hline
V505\_Per B& 0.3990 & 0.0210 & -0.1200 & 0.0300 & 0.3387 & 0.0677 & 1.2540 & 0.0072 \\ \hline
UX\_Men A& 0.3820 & 0.0290 & 0.0400 & 0.1000 & 0.5367 & 0.1073 & 1.2350 & 0.0058 \\ \hline
UX\_Men B& 0.3200 & 0.0300 & 0.0400 & 0.1000 & 0.5367 & 0.1073 & 1.1957 & 0.0072 \\ \hline
AI\_Phe A& 0.6870 & 0.0440 & -0.1400 & 0.1000 & 0.8507 & 0.1701 & 1.2336 & 0.0045 \\ \hline
AI\_Phe B& 0.6720 & 0.0430 & -0.1400 & 0.1000 & 0.8507 & 0.1701 & 1.1934 & 0.0041 \\ \hline
WZ\_Oph B& 0.4060 & 0.0290 & -0.2700 & 0.0700 & 0.6757 & 0.1351 & 1.2268 & 0.0071 \\ \hline
WZ\_Oph A& 0.4030 & 0.0290 & -0.2700 & 0.0700 & 0.6757 & 0.1351 & 1.2201 & 0.0062 \\ \hline
V 432 Aur B& 0.4290 & 0.0230 & -0.6000 & 0.0500 & 0.8507 & 0.1701 & 1.0786 & 0.0053 \\ \hline
alf Cen A& 0.1900 & 0.0080 & 0.2400 & 0.0400 & 0.8507 & 0.1701 & 1.1050 & 0.0070 \\ \hline
alf Cen B& -0.3030 & 0.0210 & 0.2400 & 0.0400 & 0.8507 & 0.1701 & 0.9340 & 0.0060 \\ \hline
NGC188\_KR\_V12 A& 0.3440 & 0.0320 & -0.1000 & 0.0900 & 1.3482 & 0.2696 & 1.1034 & 0.0074 \\ \hline
NGC188\_KR\_V12 B& 0.3050 & 0.0320 & -0.1000 & 0.0900 & 1.3482 & 0.2696 & 1.0811 & 0.0068 \\ \hline
V568\_Lyr A& 0.2580 & 0.0320 & 0.4000 & 0.1000 & 1.6973 & 0.3395 & 1.0745 & 0.0077 \\ \hline
V568\_Lyr B& -0.5150 & 0.0360 & 0.4000 & 0.1000 & 1.6973 & 0.3395 & 0.8273 & 0.0042 \\ \hline
V636 Cen A& 0.0530 & 0.0250 & -0.2000 & 0.0800 & 0.2690 & 0.0538 & 1.0518 & 0.0048 \\ \hline
V636 Cen B& -0.4130 & 0.0350 & -0.2000 & 0.0800 & 0.2690 & 0.0538 & 0.8545 & 0.0030 \\ \hline
Sun & 0.0000 & 0.0001 & 0.0000 & 0.0050 & 1.0000 & 0.0000 & 1.0000 & 0.0001 \\ \hline
 \tableline 
 \end{tabular}
 \end{table*}

Using the observational data on Table \ref{data} we perform multi-dimensional fits, using inverse problem techniques 
associated to least square methods. Such guarantees a minimum deviation between the considered model considered 
and the observational data points \citep{Menke_1989}.
The matrixial form of the equation we have to solve is: 
\begin{equation}\label{ini}
\log m^{obs}=Gd,
\end{equation}
where $m^{obs}$ is the matrix with observed masses, $G$ is the matrix that contains the information related to luminosity, metallicity, and age, 
and $d$ is the matrix of parameters that we obtain using inverse problem techniques.\\
The empirical model which we will consider is linear in the parameters and goes up to the third degree of the input parameters, as presented in equation \ref{empmod}:
\begin{equation}\label{empmod}
\log m=\sum_{i=1}^N\sum_{j=1}^3 a_{i,j}x_{i}^j
\end{equation}
{Where $a_i$ are the constant parameters, $x$ describe the observational values, the index $j$ translate the exponent of $x$. 
The index $i$ describe the cases where Log $L/L_\odot$ is considered ($i=1$), the case where metallicity is considered ($i=2$) and, finally, when $Stellar Age /Age_\odot$ is used ($i=3$). Depending on how many parameters are used on the fit: N can be equal to 1 when only luminosities are being used; 
can be equal to 2 when both luminosity and metallicity are being used; and, finally , equal to 3 when all elements are being used.\\

Because our matrix system is overdetermined we have to rewrite equation \ref{ini} to determine the parameter in $d$. 
We obtain the final equation considering the least square approach shown in the equation \ref{par}.

\begin{equation}\label{par}
d^{est}=(G^TG)^{-1}G^T\log m^{obs}
\end{equation}

To take into account the errors in the measurements we use Monte Carlo techniques, in order to estimate how these influence the final results on the fit and on the calculated mass. 
The equation \ref{par} is computed 10000 times using gaussian random simulations of data points. Each simulated data point has the standard deviation associated to its 
observational error (see Table \ref{data}). Note that when $N=1$ the Monte Carlo estimated error bars are very small, because
the errors in the observed luminosity are also small (compared to the
absolute value of brightness) and they do not significantly change the final adjustment. 
Naturally the individual errors are larger for $N=2$ and $N=3$.

{Equations \ref{rlum}, \ref{lrummet} and \ref{rlummetage} show the relation we obtain for $N=1$, with $R^2 = 0.839$,  $N=2$, with $R^2 = 0.898$, and $N=3$, with $R^2 = 0.955$, respectively:

\begin{eqnarray}
\label{rlum}
\log m^{obs}&=&0.208(\pm0.011)Log L/L_\odot\nonumber\\
&&+0.063(\pm0.007)(Log L/L_\odot)^2 \nonumber\\
&&-0.130(\pm0.048)(Log L/L_\odot)^3
\end{eqnarray}

\begin{eqnarray}
\label{lrummet}
\log m^{obs}&=&0.238(\pm0.014)Log L/L_\odot\nonumber\\
&&+0.130(\pm0.041)(Log L/L_\odot)^2\nonumber\\
&&-0.280(\pm0.081)(Log L/L_\odot)^3\nonumber\\
&&+0.044(\pm0.027)[Fe/H]\nonumber\\
&&-0.167(\pm0.075)[Fe/H]^2\nonumber\\
&&-0.116(\pm0.164)[Fe/H]^3
\end{eqnarray}

\begin{eqnarray}
\label{rlummetage}
\log m^{obs}&=&0.219(\pm0.023)Log L/L_\odot\nonumber\\
&&+0.063(\pm0.060)(Log L/L_\odot)^2\nonumber\\
&&-0.119(\pm0.112)(Log L/L_\odot)^3\nonumber\\
&&+0.079(\pm0.031)[Fe/H]\nonumber\\
&&-0.122(\pm0.119)[Fe/H]^2\nonumber\\
&&-0.145(\pm0.234)[Fe/H]^3\nonumber\\
&&+0.144(\pm0.062)Stellar Age /Age_\odot\nonumber\\
&&-0.224(\pm0.104)(Stellar Age /Age_\odot)^2\nonumber\\
&&-0.076(\pm0.045)(Stellar Age /Age_\odot)^3
\end{eqnarray}

On Figures \ref{L}, \ref{Fe_L}, and \ref{Fe_L_I} we plot the calculated mass using the parameters obtained by the fit as a function of observed mass. 
To help with the visual interpretation of the quality of the model we draw the one-to-one line on each plot.

\begin{figure}[H] 
   \centering
   \includegraphics[width=\columnwidth]{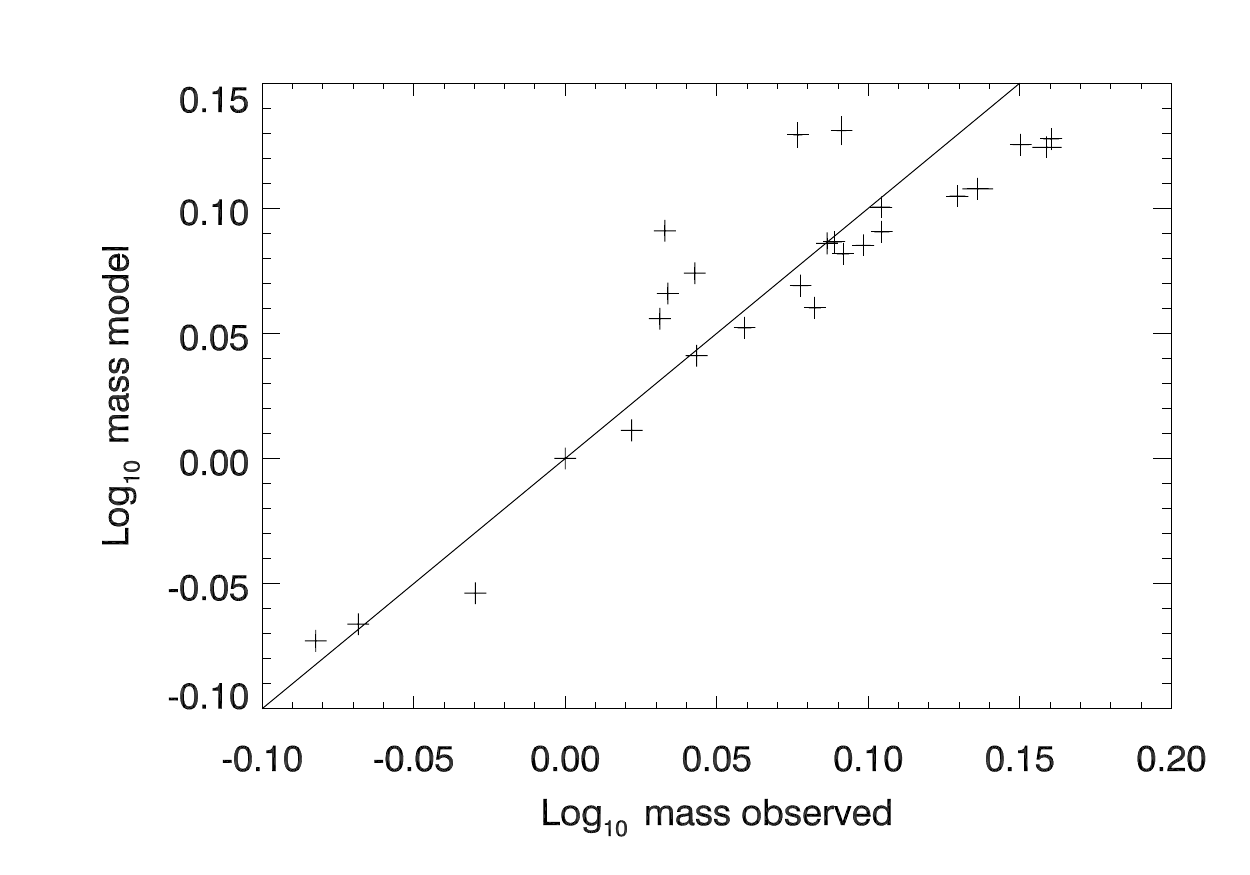} 
   \caption{Comparison between observed mass and mass obtained by the model when luminosity is considered. The one-to-one solid line is plotted to guide the eye.}
   \label{L}
\end{figure}

 \begin{figure}[H] 
   \centering
   \includegraphics[width=\columnwidth]{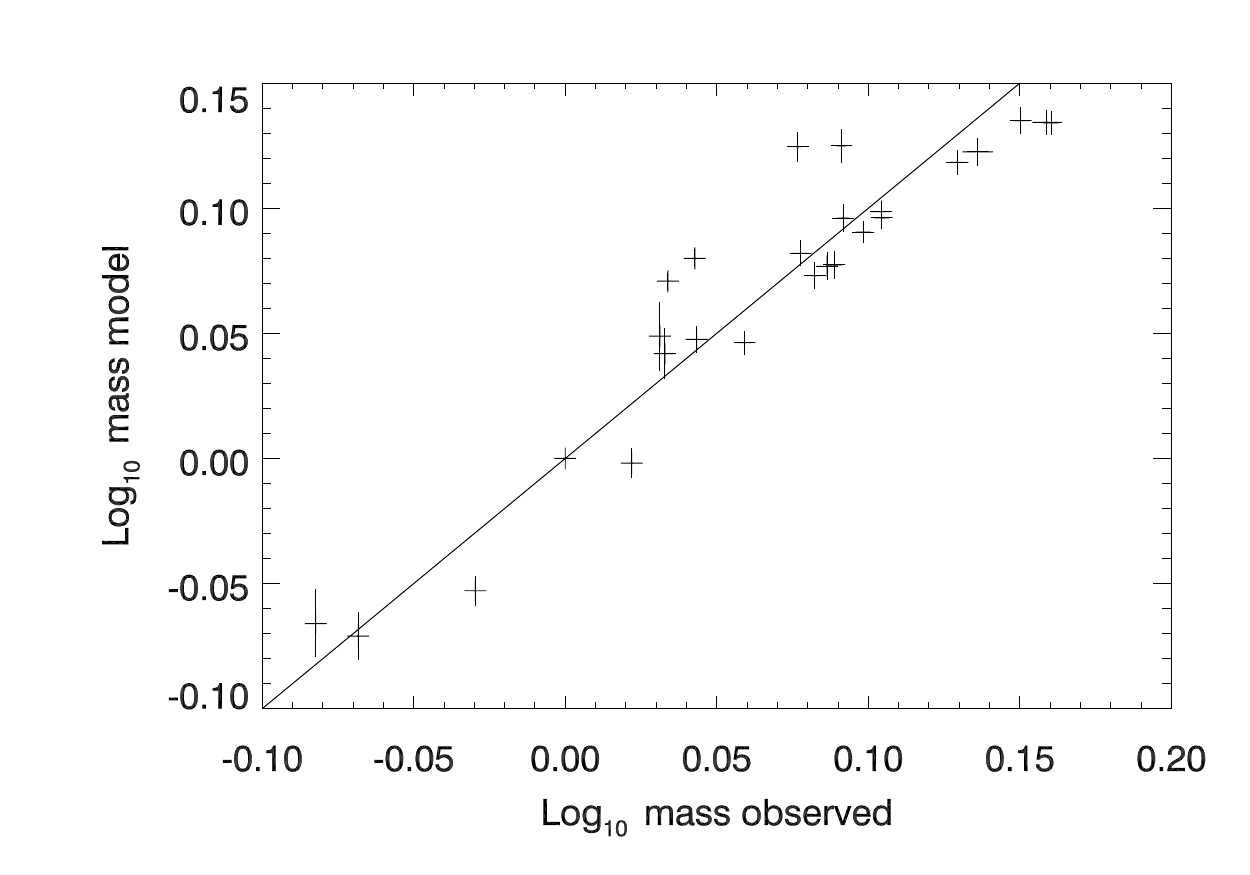} 
   \caption{Comparison between observed mass and mass obtained by the model when luminosity and metallicity are considered. 
   The one-to-one solid line is plotted to guide the eye.}
   \label{Fe_L}
\end{figure}

 \begin{figure}[H] 
   \centering
   \includegraphics[width=\columnwidth]{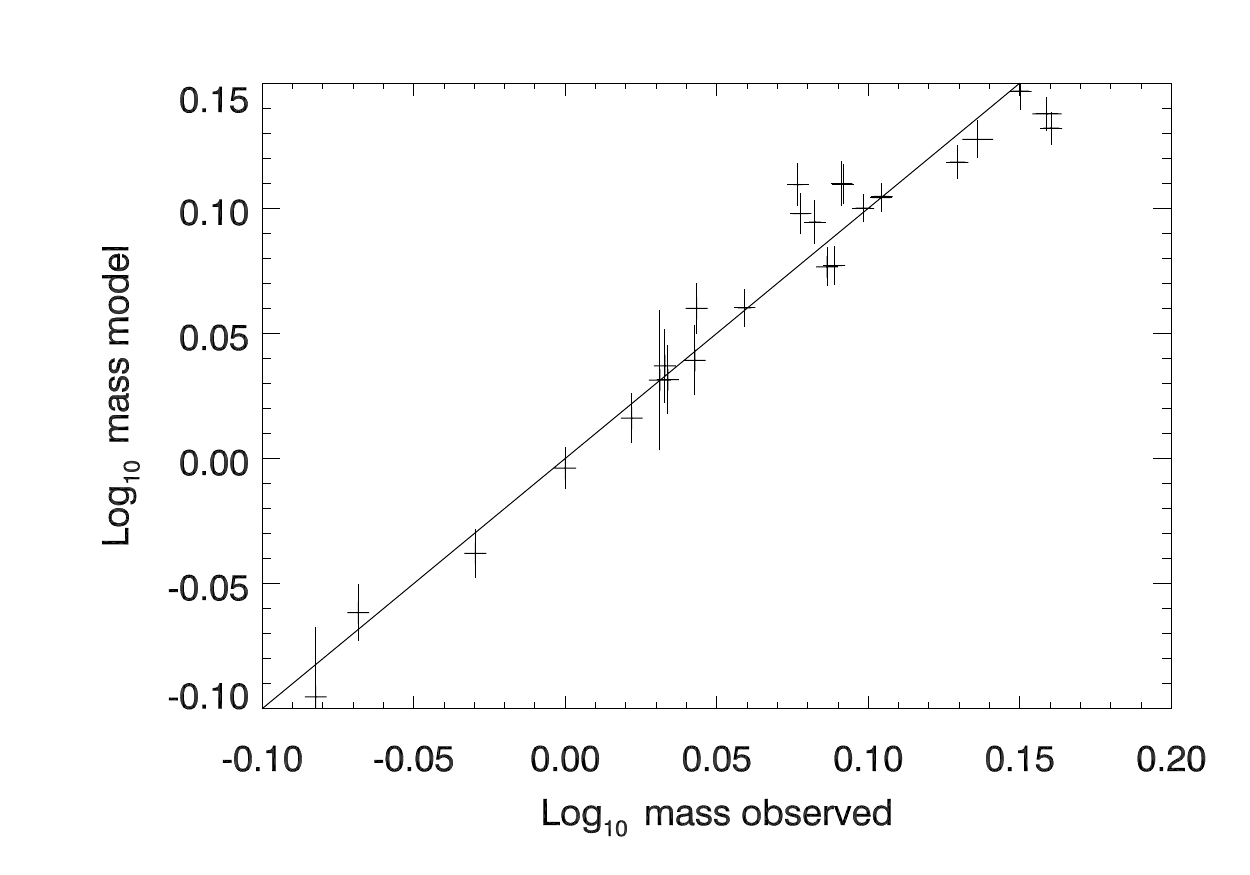} 
   \caption{Comparison between observed mass and mass obtained by the model when luminosity, metallicity, and age are considered. 
   The one-to-one solid line is plotted to guide the eye.}
   \label{Fe_L_I}
\end{figure}

\section{Discussion and conclusion}\label{dis}
In the previous section we derive three MLR: i) the classical MLR where luminosity and mass appear explicitly; ii) a MLR including the age contribution; 
and iii) the MLR including both metallicity and age contributions. 
From Figures \ref{L}, \ref{Fe_L} and \ref{Fe_L_I} we can see that our fits reproduce rather well the observed mass --- solar mass is exactly reproduced in Figures \ref{L} and \ref{Fe_L}; and for Figure \ref{Fe_L_I} we find 0.98$M_\odot$) --- sowing a natural increasing deviation for higher masses and for more evolved stars. 
However, the quality of the fit improves from Figure \ref{L} to Figure \ref{Fe_L_I}, which can be seen by the increase of the correlation coefficient from $\sim 0.84$ to $\sim 0.96$
\footnote{Note that the correlation coefficient is a non-linear parameter ranging from -1 to 1. That is, the difference between 0.8 and 0.9 is much higher than the difference between 0.1 and 0.2, for instance.}. On the other hand, our results shows (Figure \ref{dif_mass}) that if the fit uses only the luminosity, 10 stars among the total of 26 have a predicted mass outside the range $m^{obs}\pm 0.055M_\odot$ \citep[typical uncertainties for MLR kind relations][\eg]{1993AJ....106..773H}; when using both the luminosity and the metallicity, only 6 stars miss that range; and when fitting using luminosity, metallicity, and age the number of bad predictions is reduced to 3 stars. 
These results show that the inclusion of both the metallicity and the age on the MLR improves the mass predictions. 

We estimate that, on average, the non-inclusion of metallicity and age on the mass predictions through the MLR, can over/under estimate the stellar mass by 0.05$M_\odot$ (with a maximum value of about 0.15$M_\odot$)

 \begin{figure}[H] 
   \centering
   \includegraphics[width=2.9in]{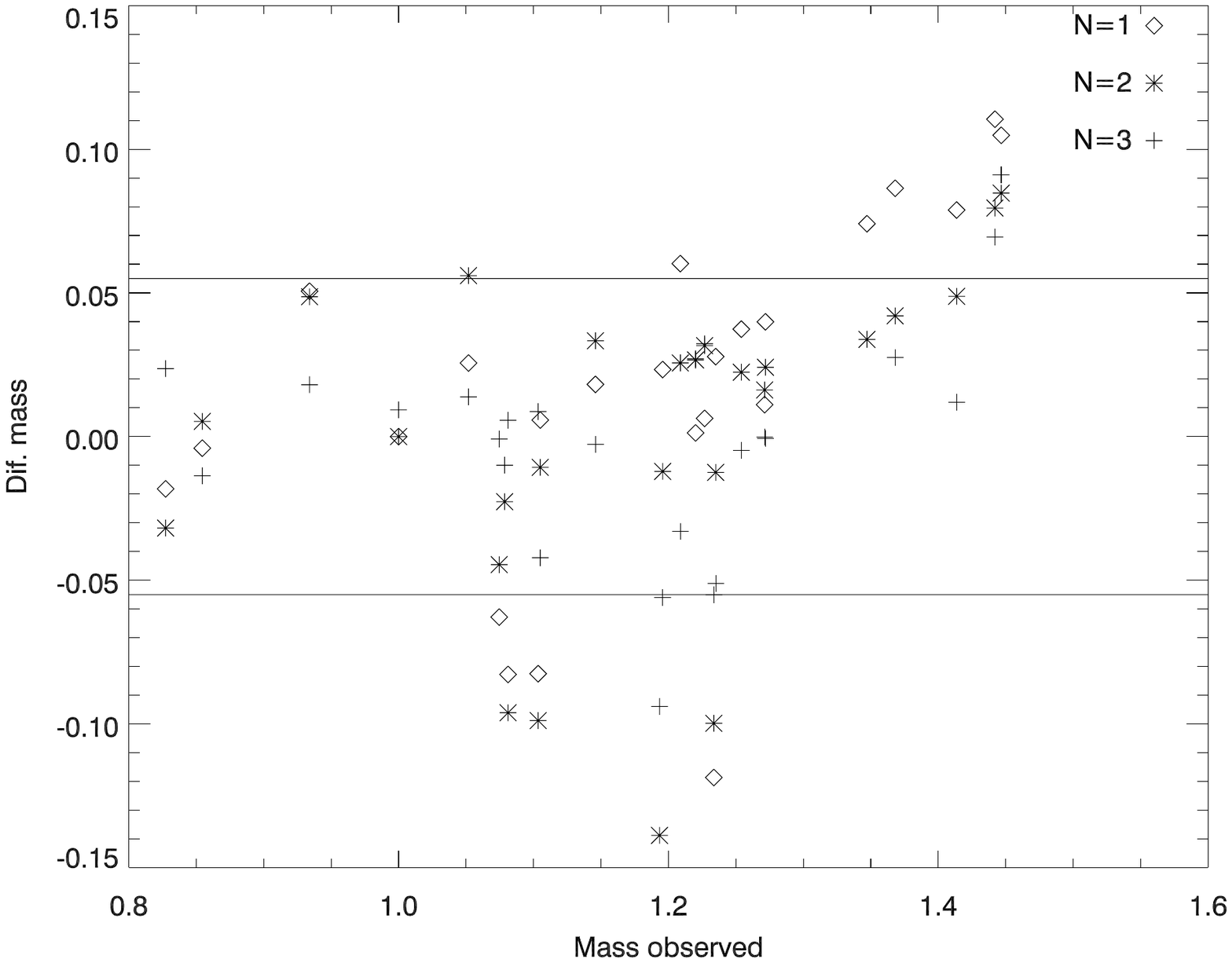} 
   \caption{The predicted mass minus the observed mass value as function of the observed mass: mass-luminosity relation (diamond), mass-luminosity-metallicity relation (star) and mass-luminosity-metallicity-age relation (cross). The horizontal line defines the $\pm 0.055M_\odot$ range.}
   \label{dif_mass}
\end{figure}

\section{Annex: Chemical composition impact on the MLR. Homological approach.}\label{annex}
According the homology approach applied to the stellar structure equations and assuming $\epsilon = \epsilon_0 \rho^\lambda T^\nu$ and $\kappa = \kappa_0 \rho^n T ^{-s}$ for the energy production rate and the opacity, respectively \citep[\cf][]{cox68}, we can write:
\begin{equation}\label{eq:r25}
4d(\ln R) - n d(\ln \rho)+ (4+s)d(\ln T) - d(\ln \emph{L}) = d(\ln
M)
\end{equation}
On the other hand re-writing the above equation in relation to Sun, we can derive:
\begin{equation}\label{eq:relacao}
\frac{L(r)}{L_\odot(r)} =
\left(\frac{\epsilon_0}{\epsilon_{\odot0}}\right)^{-\alpha}
\left(\frac{\kappa_0}{\kappa_{\odot0}}\right)^{-\beta}
\left(\frac{\mu_0}{\mu_{\odot0}}\right)^{\gamma}
\left(\frac{M}{M_\odot}\right)^{\delta}
\end{equation}
where
\begin{eqnarray*}
\alpha=\frac{3\lambda+\nu}{3\lambda+\nu-s+3n}-1\\
\beta=\frac{3\lambda+\nu}{3\lambda+\nu-s+3n}\\
\gamma=\nu-\frac{\left(\nu-s-4\right)\left(3\lambda+\nu\right)}{3\lambda+\nu-s+3n}\\
\delta=\lambda+\nu+1-\frac{\left(3\lambda+\nu\right)\left(\lambda+\nu-s+n-2\right)}
{3\lambda+\nu-s+3n}
\end{eqnarray*} 
We want this equation written explicitly as a function of the chemical composition (X, Y and Z). 
For that we consider that the main source of opacity comes from the bound-free and free-free contributions: $\kappa = \kappa_{bf} + \kappa_{ff}$, where
\begin{equation}
\kappa_{bf} \propto 4\times10^{25}Z\left(1+X\right)\rho T^{-3.5}cm^2 g^{-1}
\end{equation}
\begin{equation}
\kappa_{ff} \propto
4\times10^{22}\left(X+Y\right)\left(1+X\right)\rho T^{-3.5} cm^2
g^{-1}
\end{equation}
On the other hand assuming the pp-chain 
\begin{displaymath}
\varepsilon(pp) = \frac{2.4\times 10^4 \rho X^2}{T_9^{\frac{2}{3}}}e
^{\frac{-3.380}{T_9^{\frac{1}{3}}}}
\end{displaymath}
 and $Y \simeq 2Z + Y_p$ (where $Y_p\sim 0.25$ is the primordial helium value, \cite{tsivilev}) we can finally derive a MLR depending on metal abundance Z:
\begin{eqnarray}
\left(\frac{M}{M_\odot}\right)^{5.46}  = \left(\frac{1-3Z-Y_p}{1-3Z_\odot -
Y_p}\right)^{26.00}\times \nonumber \\
\left(\frac{(2-3Z-Y_p)(1-Z+10^3Z)}{(2-3Z_\odot-Y_p)(1-Z_\odot
+10^3Z_\odot)}\right)^{1.08}
\times \nonumber \\
\times \left(\frac{2-3Z_\odot-\frac{5}{4}Y_p}{2-3Z-\frac{5}{4}Y_p}
\right)^{-7.77}\times\left(\frac{L(r)}{L_\odot(r)}\right) \nonumber \label{eq:pp}\\
\end{eqnarray}
Therefore, using this simple approximation, we can conclude that for the solar luminosity a change of the metal abundance (Z) from 0.004 and 0.03 
(typical for the metallicity ranges of this work) can have an impact of about 0.25$M_\odot$. This value is luminosity independent for FGK main sequence stars.

%
 \section*{Acknowledgments}

We would like to thank the anonymous referee whose comments have much improved the presentation of the paper.
We also thank Nuno Peixinho for his comments on this manuscript and Alexandra Pais for the help concerning inverse problem techniques. 


 \bibliographystyle{unsrt}  
 

\end{document}